\documentclass[onecolumn,12pt]{article}
\usepackage{amssymb}

\newcommand{\alg}[1]{\mbox{$\mathfrak{#1}$}}
\newcommand{\abs}[1]{\left|#1\right|}

\newcommand{\id}{\mbox{\bf 1}}

\newcommand{\ket}[1]{\vert #1\rangle}

\newcommand{\eg}{e.\,g.,\ }
\newcommand{\ie}{i.\,e.,\ }

\newcommand{\ten}{\otimes}
\renewcommand{\H}{\cal H}
\renewcommand{\epsilon}{\varepsilon}

\begin{document}

\title{Explaining the {\em Un}observed---Why Quantum Mechanics Ain't
Only About Information}
\author{Amit Hagar\thanks{Philosophy Department, University of Delaware,
Newark, DE 19716 USA; Email: hagar@udel.edu} \and Meir
Hemmo\thanks{Philosophy
Department, University of Haifa, Haifa, Israel 31905; Email:
meir@research.haifa.ac.il}}
\maketitle
\begin{abstract}
\noindent A remarkable theorem by Clifton, Bub and Halvorson
(2003) (CBH) characterizes quantum theory in terms of
information--theoretic principles. According to Bub (2004, 2005)
the philosophical significance of the theorem is that quantum
theory should be regarded as a ``principle'' theory about
(quantum) information rather than a ``constructive'' theory about
the dynamics of quantum systems. Here we criticize Bub's principle
approach arguing that if the mathematical formalism of quantum
mechanics remains intact then there is no escape route from
solving the measurement problem by constructive theories. We
further propose a (Wigner--type) thought experiment that we argue
demonstrates that quantum mechanics on the information--theoretic
approach is incomplete.
\end{abstract}

\newpage

\section{Introduction}
Quantum information theory has by now become to a large extent a
new orthodoxy in the foundations of quantum mechanics. It is
sometimes further claimed that the information--theoretic approach
brings out the so--called ``futility'' of the long-standing debates
over the interpretations of quantum mechanics.\footnote{For such a
claim see Fuchs (2002); For a response---Hagar (2003)}
A major conceptual tool enhancing the
information--theoretic approach is the remarkable theorem by
Clifton, Bub and Halvorson (2003, CBH henceforth) according to
which quantum theory can be characterized by three
information--theoretic principles: no signaling, no broadcasting
and no (unconditionally secure) bit commitment (NO
BIT henceforth). The aim of this paper is to examine the
information--theoretic approach to quantum mechanics focusing on
Bub's (2004, 2005) recent analysis of it and some of its
implications.

On the basis of the above three principles of the CBH theorem, Bub
(2004, p. 242; see also 2005) argues for the following three theses:
\begin{quote}
\begin{enumerate}
\item A quantum theory is best understood as a theory about the
possibilities and impossibilities of information transfer as
opposed to a theory about the mechanics of nonclassical waves or
particles.\footnote{By {\em information} Bub means information in
the physical sense as measured, \eg in quantum mechanics by the von
Neumann entropy.} \item Given the three information--theoretic
constraints, any mechanical theory of quantum phenomena that
includes an account of the measuring instruments that reveal these
phenomena must be empirically equivalent to a quantum theory.
\item Assuming the information--theoretic constraints are in fact
satisfied in our world, no mechanical theory of quantum phenomena
that includes an account of measurement interactions can be
acceptable, and the appropriate aim of physics at the fundamental
level then becomes the representation and manipulation of
information.
\end{enumerate}
\end{quote}
In his recent paper in this journal, Bub (2005) depicts the
philosophical significance of the CBH theorem as analogous to
Einstein's shift from the constructive view of
theories---attributed to Lorentz and FitzGerald---towards the
principle view of theories in the context of the special theory of
relativity.\footnote{For more on the constructive
approach to special relativity see J\'anossy (1971), Bell (1976), and Brown
(2006).} His idea is that the distinction between constructive
theories and theories of principle is suitable to characterize the
difference between the information--theoretic approach and all
other interpretations of quantum theory. As the above theses show,
Bub believes that if, indeed, the three information--theoretic
principles of the CBH theorem hold in our world, then no
constructive theory for quantum phenomena is possible that yields
different predictions than those of quantum theory.

Bub's three theses have very important implications regarding the
understanding of quantum mechanics and the future direction in
which the research at the foundations of the theory will go. We
agree with Bub's second thesis (that any constructive theory
for quantum phenomena which satisfies the three
information--theoretic principles of the CBH theorem is
empirically indistinguishable from quantum theory). Nevertheless,
here we wish to examine Bub's approach and present an alternative
view on the philosophical significance of the CBH theorem and on
the role of constructive quantum mechanical theories.\footnote{We
set aside the more general issue of the aim of physics as stated
by Bub's third thesis. In this context it is interesting that
Maxwell, although accepting the distinction between the physics of
principles and the construction of models and even admitting that
\emph{in principle} indefinitely many dynamical models can explain
certain phenomena, nevertheless devoted his career almost solely
to the construction of such models (Harman 2001). Einstein
himself, after promoting in 1905 the distinction he borrowed from
Maxwell and Poincar\'e between principles and constructions,
shifted back to the constructive view and later on abandoned what
he called ``the new fashion'' which he himself helped creating
(Balashov and Janssen 2003).} As Bub (2004) himself notes (and in
accord with his second thesis) a certain constructive theory for
quantum phenomena, namely the collapse theory by Ghirardi, Rimini
and Weber (GRW) \emph{does} give different predictions from those
of quantum theory. We conjecture, following Bub (2005), that the
GRW theory violates the (above) NO BIT constraint (see below for
some discussion about this point). However, the GRW theory itself
implies that this violation is {\em compatible} with everything we
know empirically about the physical world. Roughly, the NO BIT
constraint implies the {\em unrestricted} validity of the
superposition principle, and in particular, it entails that
macroscopic massive systems can be in nonlocal entangled EPR--type
states even with respect to their spatial degrees of freedom. But
the existence of such macrostates has never been experimentally
confirmed, so we do not really know empirically whether or not
such states are physically feasible. In this paper we focus only
on collapse theories as constructive alternatives to Bub's
principle approach.\footnote{We do not address here the subtler
issue of whether or not Bohmian mechanics might be distinguished
empirically from other no collapse theories such as modal and many
worlds theories and Bub's principle approach. We agree with Bub
(2005) that Bohmian mechanics is empirically equivalent to no
collapse quantum mechanics (but compare Valentini 2002). On the
other hand, questions of theory choice depend on quite complex
factors and not only on empirical content.} We shall argue for the
following alternative theses.
\begin{quote}
\begin{enumerate}
\begin{sloppypar}\item [I.] No collapse quantum mechanical theories
(including information--theoretic approaches, if the latter are
committed to no collapse dynamics) and alternative collapse
theories, such as the GRW theory are {\em empirically}
distinguishable. This is obvious but in the present context
deserves attention.\footnote{There are some definite suggestions
of crucial tests between the GRW theory and standard quantum
mechanics which bear on the implications of the different rates of
GRW collapses and decoherence (see Adler 2005, Adler {\em et al}.
2005, Bassi {\em et al}. 2005, Hemmo and Shenker 2005).}
%\end{sloppypar}
\item [II.] Constructive quantum mechanical
theories\footnote{By this we include GRW--type collapse theories,
hidden variables theories such as Bohm's and modal theories, and
also many worlds theories.} are necessary to solve the measurement
problem (given the current mathemtaical formalism of quantum
theory). \item [III.] Quantum mechanics under the
information--theoretic interpretation is {\em incomplete}, and
moreover, it leads to inconsistent predictions of the
(statistical) outcomes of (some) measurements.
\end{sloppypar}\end{enumerate}\end{quote}
As to Thesis II, we shall also argue that interpreting quantum
mechanics as a principle theory is {\em not} the right
epistemological stance given the theoretical basis of quantum
mechanics, and that information--theoretic interpretations of
quantum mechanics as a principle theory (see Bub 2004, 2005) don't
warrant abandoning alternative constructive dynamical theories, in
particular theories which differ empirically from no collapse
quantum mechanics. We shall further argue that the lesson one
should take from the CBH theorem lies not only in the quantum
phenomena that are captured by its three information--theoretic
conditions, but also in the predictions of quantum mechanics,
given the general validity of the CBH conditions, which up to now
remain {\em un}observed. As to Thesis III (above), we shall argue
by explicit construction that the information—theoretic approach
must be supplemented by further principles over and above those
suggested by the CBH theorem. Also, we shall argue that the notion
of quantum information cannot be taken as a primitive but rather
requires (as in any other quantum theory) a quantum mechanical
analysis of measurement of the kind suggested by constructive
theories. Finally, we shall argue that a {\em complete} and
consistent information—-theoretic approach is bound to rely on
{\em constructive} models, mainly because the measurement problem
in quantum mechanics can only be solved by constructive theories;
it cannot be otherwise bypassed by theories that treat
measurements as ``black boxes''.

The paper is structured as follows. We briefly review in Section
\ref{CBH} the CBH theorem and the purported philosophical
significance Bub attaches to it. In Section \ref{unob} we explain
how the GRW collapse theory bears on the CBH
information--theoretic principles, focusing (Section \ref{grw}) in
particular on the NO BIT principle that Bub sees as constraining
any constructive model for quantum phenomena. In Section \ref{cp}
we argue for thesis II above, and in particular that the issue at stake
is different explanations of as yet {\em un\/}observed quantum
predictions. In Section \ref{incon}, we focus on the empirical
inequivalence between the constructive GRW theory and
information--theoretic approaches: we first challenge the latter
with a thought experiment that establishes our thesis III above
(Section \ref{exp}); and we consider various possible replies to
our argument in Section \ref{res}. Finally, in Section \ref{anc}
we consider another often stated argument against collapse
theories and explain why this argument is unacceptable.

\section{The CBH Theorem and Its Philosophical Significance}\label{CBH}
The question raised by CBH is whether we can deduce the kinematic
aspects of the quantum--theoretic description of physical systems
from the assumption that we live in a world in which there are
certain constraints on the acquisition, representation, and
communication of information. CBH answered this question
positively, supplying three information--theoretic principles
(so-called three {\em no-go's}) that are supposed to filter out
the algebraic structure of operators and states that characterize
(what they take to be) quantum theory from the more basic structure
of $C^*$-algebras.

The first principle, called {\em no signaling}, prohibits
superluminal transfer of information between spacelike separated
systems by carrying out measurements on one of them. In other
words, no signaling says that measurements (and in fact any
physical manipulation) confined to a remote system cannot possibly
change the statistics of the outcomes of measurements that might
be carried out on the local system. If Alice and Bob are two
physically distinct systems,\footnote{Consider a composite quantum
system A+B, consisting of two subsystems, A and B. For simplicity,
assume the systems are identical, so their $C^{*}$--algebras
$\alg{A}$ and $\alg{B}$ are isomorphic. The observables of the
component systems A and B are represented by the self-adjoint
elements of $\alg{A}$ and $\alg{B}$, respectively. Let
$\alg{A}\vee\alg{B}$ denote the $C^{*}$--algebra generated by
$\alg{A}$ and $\alg{B}$. To capture the idea that A and B are
\textit{physically distinct} systems, we assume (as a necessary
condition) that any state of $\alg{A}$ is compatible with any
state of $\alg{B}$, i.e., for any state $\rho _{A}$ of $\alg{A}$
and $\rho _{B}$ of $\alg{B}$, there is a state $\rho$ of
$\alg{A}\vee \alg{B}$ such that $\rho |_{\alg{A}}=\rho _{A}$ and
$\rho |_{\alg{B}}=\rho _{B}$.} then when both perform local
measurements, Alice's measurements can have no influence on the
statistics of the outcomes of Bob's measurements, and conversely.
This result follows from the {\em no signaling} theorem in quantum
mechanics according to which local measurements on a system
$\alpha$ have no effect whatsoever on the reduced state of a
remote system $\beta$ no matter what the quantum state of
$\alpha+\beta$ is. (see Ghirardi {\em et al}. 1980).

The second principle, called {\em no broadcasting}, prohibits
perfectly broadcasting the information contained in an unknown
physical state.\footnote{In fact, for pure states, broadcasting
reduces to cloning. In cloning, a ready state $\sigma$ of a system
B and the state to be cloned $\rho$ of system A are transformed
into two copies of $\rho$. In broadcasting, a ready state $\sigma$
of B and the state to be broadcast $\rho$ of A are transformed to
a new state $\omega$ of A+B, where the marginal states of $\omega$
with respect to both A and B are $\rho$.} No broadcasting ensures
that the individual algebras $\alg{A}$ and $\alg{B}$ of the two
distinct physical systems are noncommutative. As CBH show,
broadcasting and cloning are always possible for classical
systems, i.e., in a commutative $C^*$--algebra there is a universal
broadcasting map that clones any pair of input pure states and
broadcasts any pair of input mixed states. Conversely, they show
that if any two states can be (perfectly) broadcast, then any two
pure states can be cloned; and if two pure states of a
$C^{*}$-algebra can be cloned, then they must be orthogonal. So,
if any two states can be broadcast, then all pure states are
orthogonal, from which it follows that the algebra is commutative.
In elementary quantum mechanics, on the other hand, neither
cloning nor broadcasting is possible in general.

These two principles capture two well known features of quantum
theory: for a composite system A+B, the no signaling constraint
entails that the $C^{*}$--algebras $\alg{A}$ and $\alg{B}$, whose
self-adjoint elements represent the observables of A and B,
commute with each other (this feature is sometimes called
`micro-causality'); and the no broadcasting constraint entails
that each of the algebras $\alg{A}$ and $\alg{B}$ is
noncommutative. The quantum mechanical phenomenon of interference
is the physical manifestation of the noncommutativity of quantum
observables or, equivalently, the superposition of quantum states.

The third NO BIT principle prohibits communicating information in
a way that implements a given `bit commitment' with {\em
unconditional} security.\footnote{Bit commitment is a
cryptographic protocol in which, say, Alice sends an encoded bit
to Bob as a record of her commitment to either $0$ or $1$, which
allows Bob to ascertain Alice's bit commitment later (only with
further information supplied by Alice) so as to make sure that her
initial commitment hasn't changed. In classical information theory
unconditionally secure bit commitment is always possible in
principle.} In quantum mechanics Alice may send Bob, as a warrant
of her bit commitment, one of {\em two} mixtures associated with
the same density operator (where the mixtures correspond to
alternative commitments). However, Alice may prepare in advance a
suitable entangled state, where the reduced density operator for
Bob is the same as that of the mixture she sent him. In this case
Alice would be able to steer Bob's system nonlocally into either
one of the two mixtures (where Bob cannot be aware of this). So if
there are no restrictions on the entangled states that Alice may
prepare, Alice can always cheat Bob by pretending to have a secure
bit commitment.

The NO BIT constraint prohibits unconditionally secure bit
commitment by stipulating that there are no restrictions on the
preparation and stability of entangled nonlocal states. Note that
the structure that the first two principles filter out from the
general $C^*$--algebra still includes noncommutative theories
which are compatible with unconditionally secure bit commitment.
In such theories, it might be, for example, that although some
nonlocal entangled states (\ie which permit remote steering) are
physically possible, they turn out to be highly unstable (over
time) by the dynamical equations of motion (say, given a
non-unitary dynamics) and therefore not feasible.\footnote{As
noted by Bub, such a possibility in which an entangled state of a
composite system quickly decays to a mixture as soon as the
component systems spatially separate was raised by Schr\"odinger
in 1936.} So one has to stipulate the feasibility or dynamical
{\em stability} of such states, and this is what the NO BIT
constraint does over and above the other two conditions of no
signaling and no broadcasting.\footnote{Timpson (2004, Ch. 9)
argues that the NO BIT condition follows logically from no
signaling and no broadcasting, and is therefore redundant.
However, in theories with non--unitary dynamics (\eg
Schr\"odinger--type theories) some pure states in the Hilbert
space of a system (\eg superpositions of position states
corresponding to spatially separated systems), although
permissible, turn out to be dynamically unstable in the sense that
they decay extremely quickly by the equations of motion into mixed
states that correspond to classical mixtures. In this sense the NO
BIT condition in a Schr\"odinger--type theory (that satisfies no
signaling and no broadcasting) might turn out to be unstably
sustained for some states. An example of a Schr\"odinger--type
theory of this kind is the GRW theory. We shall argue for our
conjecture that the GRW theory does violate the NO BIT condition
in the above dynamical sense and that this is the sense relevant
to Bub's analysis in Section \ref{cp}, and footnote 19.}

Taking stock, the content of the CBH theorem, according to Bub
(2005), is this:

\begin{quote}
$\ldots$ [Q]uantum theories---theories where (i) the observables
of the theory are represented by the self-adjoint operators in a
noncommutative $C^{*}$--algebra (but the algebras of observables
of distinct systems commute), (ii) the states of the theory are
represented by $C^{*}$--algebraic states (positive normalized
linear functionals on the $C^{*}$--algebra), and spacelike
separated systems can be prepared in entangled states that allow
remote steering, and (iii) dynamical changes are represented by
completely positive linear maps---are characterized by the three
information--theoretic `no-go's': no superluminal communication of
information via measurement, no (perfect) broadcasting, and no
(unconditionally secure) bit commitment.
\end{quote}

In order to flesh out the philosophical significance of the CBH
theorem, Bub (2005) makes use of the famous distinction between
theories of principle and constructive theories. According to this
distinction (which is attributed usually to Einstein although it
already appears in the writings of Maxwell and Poincar\'e),
\begin{quote}
[constructive theories] attempt to build up a picture of the more
complex phenomena out of the materials of the relatively simple
formal scheme from which they start out. Thus the kinetic theory
of gases seeks to reduce mechanical, thermal and diffusional
processes to the movement of molecules---i.e., to build them up of
the hypothesis of molecular motion. [Principle theories, on the
other hand,] $\ldots$employ the analytic, not the synthetic
method. The elements which form their basis and starting point are
not hypothetically construed but empirically discovered ones,
general characteristics of natural processes, principles that give
rise to mathematically formulated criteria which the separate
processes or the theoretical representations of them have to
satisfy. Thus the science of thermodynamics seeks by analytical
means to deduce necessary conditions which separate events have to
satisfy, from the universally experienced fact that perpetual
motion is impossible  (Einstein 1919).
\end{quote}

In his analysis of quantum mechanics as a principle theory, Bub
appeals to two different historical analogies where scientific
progress has been clearly achieved. In his (2004) he considers the
transition from the constructive ether-theory of
Lorentz--FitzGerald to the abstract geometric formalism of
Minkowski's spacetime and argues that the transition was only made
possible by Einstein's principle approach to special relativity.
And in (2005) Bub focuses on the transition (in the `opposite'
direction) from thermodynamics (as a sort of a principle theory)
to the constructive theory of statistical mechanics (in the
special case of the kinetic-molecular theory). In both cases Bub's
historical analysis seems to be plausible (but these issues are
under dispute; compare Brown 2003, 2006), but we are doubtful
as to the conclusions he draws about quantum mechanics. Bub argues that
the CBH theorem plays the same role in a principle approach to
quantum mechanics as the one played by Einstein's principle
approach to relativity theory. Focusing on Bohmian mechanics as a
constructive mechanical model of quantum mechanics, Bub's argument
consists of essentially three elements: First, in special
relativity the structure of spacetime is understood in terms of a
new primitive---\ie a {\em field\/}---which is not reducible to
mechanical motion (\eg of particles relative to the ether as in
the Lorentz theory). Similarly, in quantum mechanics, the
algebraic structure of observables is to be understood in terms of a new
primitive, \ie {\em quantum information}, not reducible to the
behavior of mechanical systems (\eg particle trajectories in
configuration space). Second, in both cases the principle
approaches are simpler and more fruitful. In the case of quantum
theory the CBH theorem is taken to explain {\em away} (using the
above constrains on information flow) some problematic notions in
Bohmian mechanics such as sourceless fields that guide the
trajectories of particles in configuration space (that sometimes
even result in \eg surreal trajectories and contextual
probabilities).\footnote{Here we try to flesh out Bub's preference
for a principle approach to quantum mechanics over a constructive
theory such as Bohm's. Obviously, supporters of Bohm's theory
judge surreal trajectories and contextual probabilities as {\em
un}problematic.} At the same time the information--—theoretic
approach brings out new implications of quantum mechanics such as
the use of entanglement as a possible new physical resource for
quantum computation. Third, and most crucially in the present
context, the constructive mechanical alternatives to quantum
theory are {\em empirically} indistinguishable from the principle
theory.

The point of Bub's second analogy (\ie the transition from
thermodynamics to the kinetic-molecular theory) is precisely to
bring out the immense importance of empirical distinguishability
in theory choice. Here the argument is that the kinetic-molecular
theory would not be regarded seriously as an alternative
constructive model for thermodynamics if (contrary to fact) it had
no new empirical predictions that differ from those of
thermodynamics, and if those predictions were not experimentally
confirmed (\eg Einstein's prediction of fluctuations in Brownian
motion). By contrast, in the case of Bohmian mechanics it is
provable that (i) once the distribution of the particles'
positions is given by the square of the amplitude of the
wavefunction at each point (\ie by Born-like probabilities) this
distribution is preserved at all later times by the dynamical
equations of motion (see D\"urr, Goldstein and Zanghi 1992); and
(ii) given the Born distribution, Bohm's theory is empirically
equivalent to quantum mechanics. This means, in the
information--theoretic approach, that Bohm's theory must be
equivalent to quantum theory if the CBH constrains on the
information flow were satisfied even once in the past (since,
roughly, in a no collapse theory these constrains hold if and only
if the Born probability distribution holds). Consequently, Bohm's
theory, quite unlike the case of Brownian motion, can yield no
predictions of `fluctuations' that deviate from the predictions of
quantum mechanics. And therefore, given Bub's arguments above, the
rational epistemological stance, is to reject it, and prefer the
principle information--theoretic approach.

In what follows we question Bub's reading of the present state of
affairs in quantum mechanics. Although we largely agree with Bub's
analysis (sketched above) of the features of alternative hidden
variables theories, we think that the analysis doesn't capture all
the relevant aspects related to theory choice in the case of the
present state of quantum mechanics. In particular, there are two
crucial points that seem to us not appropriately addressed in
Bub's analysis. First, there are {\em other} constructive quantum
mechanical theories (in particular the GRW {\em collapse} theory)
which generally violate the NO BIT constraint (as a dynamically
stable constraint; see Section \ref{cp}). The GRW theory differs in its
empirical predictions from quantum theory, while it is perfectly
compatible with our experience so far. By Bub's own standards (see
Bub 2005, Sec. 4), therefore, the GRW theory is acceptable as an
alternative constructive theory. But adhering to Bub's principle
approach would result in loosing sight of theories like the GRW
theory on what seems like an {\em a--priori} rather than an
empirical basis. Second, information--theoretic approaches (\ie
both Bub's principle approach and the Bayesian approach) are {\em
incomplete}, and as we said (in our thesis (III) above) need be
supplemented by further principles that are quite hard to justify
(see Section \ref{incon}).

On the basis of these two points we now proceed to argue for our
thesis (II), namely that constructive theories are necessary to solve
the measurement problem in standard quantum mechanics, and that at least
partially the issue at stake lies not in the information--theoretic
description of the \emph{observed} quantum
phenomena, but rather in the explanation of the predictions of
quantum theory which up to now remain \emph{un\/}observed.

\section{Explaining the {\em{Un\/}}observed}\label{unob}
Standard no collapse quantum theory predicts the unrestrictive
existence of superpositions in spatially separated entangled states.
This is tantamount to saying, using the CBH theorem, that {\em ex
hypothesis} the NO BIT principle holds in our world. But then, given
this hypothesis, one question that arises naturally is: why do
superpositions remain \emph{un\/}observed in macroscopic
massive physical systems?

This question is in fact a variant on the so-called measurement
problem in the quantum theory of measurement. In this context the problem
arises as a straightforward consequence of applying the Schr\"odinger linear and
deterministic dynamics to the measurement interaction. As is well known
the Schr\"odinger dynamics results for a generic measurement in a
superposition of the form
\begin{equation}
 \ket{\Psi}=\sum_i\mu_i\ket{\psi_i}\ten\ket{\varphi_i},
 \label{eq:sup}
\end{equation}
where the kets $\ket{\psi_i}$ represent some suitably defined
pointer states of the measuring apparatus (typically, the
$\ket{\psi_i}$ are eigenstates of the pointer position), and the
$\ket{\varphi_i}$  are some states of the system. The problem is
that in states of the form (\ref{eq:sup}) the measurement has no
definite outcome (except in the special case where all but one of
the $\mu_i$ are zero), since the final (reduced) state of the
apparatus cannot in general be described in terms of an ensemble
of systems in a classical mixture (in which the $\abs{\mu_i}^2$
represent the probabilities for each $\ket{\psi_i}$ to actually
occur).

The information--theoretic approach addresses this problem by
appealing to models of decoherence in which the interaction of
relatively massive systems with their environment brings about a
so-called {\em effective} collapse onto the eigenstates of some
preferred observables (typically, position).\footnote{For standard
models of decoherence, see Joos {\em et al.} (2003) and references
therein.} According to this approach (called by Bub 1997 some
years ago the `new orthodoxy') macroscopic entangled states in
position exist, but for all practical purposes they are unobserved
because we have no control over the states of the environment, so
that the reduced state of a decohering system is practically
indistinguishable from a classical mixture. That this is no
solution to the measurement problem was argued in the past by
many,\footnote{E.g., Bell (1990), and recently Adler (2003).}
including Bub himself in the context of (constructive) hidden variables
theories.\footnote{As Bub puts it in his (2000, 90--91): the fact
that the `effective' quantum state---an improper mixture described
by the reduced density operator (obtained by tracing out the
degrees of freedom of the environment)---is diagonal with respect
to properties associated with some pointer basis ``not only fails
to account for the occurrence of just one of these [properties]
but is actually inconsistent with such occurrence'', since taking
into account the environment gives us back the pure state from
which the mixture was derived, and this state is inconsistent with
the occurrence of events associated with definite properties.}
However, Bub seems to think that the objection does not apply to
his own principle information--theoretic approach. We disagree and
will argue for this in Section \ref{incon}. But before doing so,
we wish to consider here the constructive collapse theory by GRW
that gives a clear and distinct solution to the measurement
problem and explains the \emph{un\/}observability of macroscopic
spatial superpositions in the most straightforward way.

\subsection{The GRW theory}\label{grw}
The GRW theory (formulated for non--relativistic quantum
mechanics) explains the {\em un\/}observability of some
macroscopic superpositions of {\em position} states by modifying
the Schr\"odinger linear dynamics in such a way that given the new
dynamics such superpositions are overwhelmingly likely to collapse
at every moment of time, and in this sense they are highly
unstable. The Schr\"odinger equation is changed by adding to it a
non-linear and stochastic term that induces the so-called {\em
jump} or collapse of the wavefunction. The jump is supposed to
occur on occasion in position space and its postulated frequency
is proportional roughly, to the mass density of the system (or in
Bell's (1987) model on the number of particles described by the
wavefunction). For our purposes it is enough to sketch Bell's
(1987) version of the elementary and non--relativistic theory of
GRW. This goes roughly as follows.

Consider the quantum mechanical wavefunction of a composite system consisting of
$N$ particles:
\begin{equation}
\psi(t, {\bf r}_1, {\bf r}_2,...,{\bf r}_N). \label{eq:psi}
\end{equation}
The time evolution of the wavefunction usually (at almost all
times) satisfies the deterministic Schr\"odinger equation. But
sometimes {\em at random} the wavefunction collapses or {\em jumps})
onto a wavefunction $\psi_\ell$ localized in position of the (normalized)
form
\begin{equation}
 \psi_\ell = \frac{j({\bf x}-{\bf r}_n)
    \;\psi(t, {\bf r}_1, {\bf r}_2,...,{\bf r}_N)}
      {R_n({\bf x})},
\label{eq:loc}
\end{equation}
where ${\bf r}_n$ in the jump factor $j({\bf x}-{\bf r}_n)$ (which is
normalized) is randomly chosen from the arguments ${\bf r}_1,...,{\bf r}_n$ of
the wavefunction immediately before the jump, and $R_n({\bf x})$ is a
suitable renormalization term. For $j$, GRW suggest the Gaussian:
\begin{equation}
 j({\bf x})=K\; {\rm exp} (-{\bf x}^2/2\Delta^2),
\label{eq:gau}
\end{equation}
where the width $\Delta$ of the Gaussian is supposed to be a new
constant of nature: $\Delta\approx 10^{-5}{\rm cm}$.

Probabilities enter the theory twice. First, the probability
that the {\em collapsed} wavefunction $\psi_\ell$ after a jump is
centered around the point ${\bf x}$ is given by
\begin{equation}
{\rm d}^3{\bf x}\abs{R_n({\bf x})}^2. \label{eq:born}
\end{equation}
This probability distribution, as can be seen, is proportional to
the standard quantum mechanical probability given by the Born rule
for a position measurement on a system with the wavefunction
$\psi(t, {\bf r})$ just prior to the jump. Second, the probability
in a unit time interval for a GRW jump is
\begin{equation}
  \frac{N}{\tau},
 \label{eq:ntau}
\end{equation}
where $N$ is the number of arguments in the wavefunction (\ie in
Bell's model it may be interpreted as the number of particles),
and $\tau$ is, again, a new constant of nature ($\tau\approx
10^{15}\,{\rm sec}\approx 10^8\; {\rm year}$). Note that the
expression (\ref{eq:ntau}) does not depend on the quantum wave
function, but only on $N$. This is essentially the whole theory.

\begin{sloppypar}
For microscopic systems GRW collapses have
extremely low probability to occur, and the quantum mechanical
Schr\"odinger equation turns out to be effectively true at almost
all times in just the way that no collapse quantum mechanics
predicts (and experiment confirms). However, for massive
macroscopic systems (\eg for systems with $10^{23}$ particles) the
GRW collapses are highly probable at all times. In measurement
situations the GRW theory implies that superpositions of
macroscopically distinguished pointer states of the form
(\ref{eq:sup}) collapse with extremely high probability onto the
localized states $\ket{\psi_i}$ on time scales that are much
faster than measurement times. In particular, the probability that
the wavefunction of the composite of system plus apparatus will
stay in the superposition (\ref{eq:sup}) for more than a fraction
of a second (\eg by the time the measurement is complete) vanishes
exponentially. So the GRW jumps reduce wavefunctions of the form
(\ref{eq:sup}) to one of the components
$\ket{\psi_i}\ten\ket{\varphi_i}$, in which the pointer is in the
{\em localized} state $\ket{\psi_i}$, where the probability for a
collapse onto the $i$-th term (see equation (\ref{eq:born})) is
given as usual by the squared amplitude $\abs{\mu_i}^2$. This
means that in a sequence of quantum mechanical measurements the
GRW jumps result in definite outcomes with frequencies that are
(approximately) equal to the Born-rule probabilities
$\abs{\mu_i}^2$. The measurement problem is solved by this construction
as long as measurements involve a macroscopic recording of the measurement
outcome in {\em position} (\eg a moving pointer of a measuring
device, particles impinging on macroscopically separated regions of
a computer screen, etc.). Note that, and this is important for our
discussion below, the GRW jumps are designed to be extremely
effective {\em only} for macroscopic superpositions of {\em
position} states (and to any other states that are {\em coupled}
to positions), but {\em not} to arbitrary superpositions.
\end{sloppypar}

There are well known physical weaknesses in the GRW theory. In the
non--relativistic case the problem seems to be how to avoid the
accumulated violations of conservation of energy induced by the
jumps.\footnote{See Ghirardi (2000) and references therein; Some
progress is reported in Bassi \emph{et al\/}. (2005b)}. But
perhaps the main problem is to write down a relativistic
formulation of the GRW theory. Although GRW give the same
predictions as standard quantum mechanics with respect to the
nonlocal correlations in EPR--Bell--type experiments, the problem
is that it is not clear whether in a relativistic version of the
GRW dynamics the jumps could be made
Lorentz--covariant.\footnote{But see Ghirardi (2000) and Myrvold
(2002).} Also, it has been argued against GRW that the theory
appears to be strongly {\em ad hoc} as it allows free adjustments of
constants in order to be in agreement with experiments. Although we
believe that these complaints are largely unjustified, here
we set this issue aside, because we are not
concerned with the GRW theory {\em per se\/}, nor with its
comparison with other constructive theories. Rather, our focus is
whether Bub's principle approach which seems to {\em rule out} the
GRW theory on the basis of information--theoretic constraints is
acceptable.

\subsection{Constructing the principles}\label{cp}
How do the information--theoretic constraints of the CBH theorem
bear on the GRW theory? The first thing to say is that the GRW
theory does not violate the first two constraints (i.e., no
signaling and no broadcasting) because of its stochastic dynamics
(see, \eg Gisin 1989). But, we conjecture (following Bub 2005)
that the GRW theory does violate the NO BIT principle in the
following sense. The GRW dynamics is devised in such a way that
superpositions of macroscopically distinguishable {\em position}
states decay extremely quickly by the equations of motion into
the corresponding classical-like mixtures.
Obviously, no superselection rules are
introduced by the GRW theory. That is, all the usual quantum
mechanical observables are permissible in the theory. And so {\em
any} pure quantum state of any system and {\em any} entangled and
nonlocal state (say, of the EPR--Bell--type) in the Hilbert space
of a composite system is permissible too. In particular, one may
prepare, for example, a superposition of highly entangled spin
states even for a macroscopic system that will be quite stable
over a significant time interval (\eg as in the spin--echo
experiments, where the spins get, as it were, in and out of
macroscopically entangled states during an appreciable time
interval; see Hemmo and Shenker 2005). Such states are perfectly
compatible with the GRW theory and might even be quite stable over
time. Moreover, two massive molecules can be made to enter with
certainty a quantum state in which their position degrees of
freedom will be highly entangled as in EPR--type states. But,
given the GRW dynamics, states of this latter sort would be highly
unstable due to the extremely fast rates of the GRW jumps in {\em
position\/}. More generally, it is an empirical prediction of the
GRW theory that superpositions of entangled position states of
spatially separated systems are highly unstable if the systems are
massive enough. Hence, the theory predicts that
observable effects of such superpositions are hard to come by,
and under normal circumstances, remain {\em un}observed.

Therefore, we conjecture, following Bub (2005), that in the GRW
theory an unconditionally secure bit commitment could always be
made possible in principle (modulo the complexity of the
encryption code) via a set up that requires Alice to encode her
bit commitment in the position state of a massive enough system
(relative to the GRW parameters that fix the probability for
spontaneous localization). The NO BIT constraint is violated in
the GRW theory in the sense that spatially separated
macro--systems cannot be stably entangled in their positions. That
is, in the GRW theory it is {\em not true} that for {\em any}
mixed state that Alice and Bob can in principle prepare by
following some (bit commitment) protocol, there exists a
corresponding nonlocal entangled state that can be physically and
{\em stably} occupied by Alice's and Bob's systems. That is, it is
a straightforward consequence of the GRW dynamics that some
nonlocal entangled states are highly unstable, in the sense that
these states---although permissible in principle---decay extremely
quickly to mixed states that essentially correspond to classical
mixtures). Thus we take the NO BIT condition (following
Bub 2005) as a {\em dynamical} condition on the stability of
quantum states, hence as filtering alternative {\em quantum}
Schr\"odinger--type theories such as the GRW
theory.\footnote{Timposn (2004, ch. 9) overlooks this point. For the
purpose of filtering out only {\em classical} theories, the NO BIT
condition is redundant, since it is implied (as Timpson argues) by no
signaling and no broadcasting. But to rule out GRW--type theories
further dynamical conditions are required. For this purpose the NO
BIT condition seems to be necessary and (together with the other two
conditions) sufficient. Perhaps, it could be narrowed down and translated
into an empirical constraint (say, in the form of a Bell-like inequality)
for each collapse dynamics.}
This, of course, is not sufficient to entail that
in the GRW theory a secure bit commitment protocol is feasible,
but the security of Alice's protocol will depend {\em only} on the
computational complexity of the encryption code.

However, the introduction of the unconditional NO BIT constraint
into quantum mechanics implicitly presupposes that as a matter of
fact the entangled states shared between Alice and Bob are states
of \emph{micro\/}systems, say, over spin degrees of freedom. For
such states we have ample experimental confirmation that the NO
BIT constraint is, indeed, satisfied. But for such states
\emph{also} the GRW theory satisfies the NO BIT constraint with
extremely high probability. It violates the NO BIT constraint in
the above sense only with respect to superpositions of
macroscopically distinguishable position states. But for such
\emph{macro\/}states also standard no collapse quantum theory
predicts an \emph{effective} violation of the NO BIT constraint in
the same sense due to environmental decoherence (\ie such states
effectively collapse also in the standard theory as all models of
environmentally induced decoherence show; see \eg Zurek 1991, Joos
{\em et al.} 2003).

So practically, no collapse quantum mechanics and the GRW theory
agree on the NO BIT constraint for all cases in which there are
good empirical reasons to believe it is true. And the two theories
seem to disagree about the NO BIT constraint only with respect to
those macro superpositions about which we do not know whether or
not they in fact exist in our world. But of course this is no
surprise since this disagreement is located precisely where the
two theories differ in their empirical predictions. So we are back
to square one! The CBH theorem therefore provides a clear cut
principle that distinguishes between the empirical predictions of
theories of genuine collapse of the GRW type and theories of
effective collapse (as in models of environmental decoherence).
But given what we know empirically about the world, there seem to
be no grounds for adopting the NO BIT principle as an
unrestrictive constraint on theory choice.\footnote{Note that the
point made here is quite general. It would be applicable to
\textit{any} information--theoretic characterization of the
mathematical structure of quantum theory (\eg Spekkens 2004) and
to any constructive alternative of it, provided the latter is
empirically well confirmed and {\em not} empirically equivalent to
standard quantum mechanics.}

\section{Inconsistent Predictions?}\label{incon}
We now proceed to argue for our thesis III, namely that
information--theoretic approaches to quantum mechanics are {\em
incomplete} and need be supplemented by further axioms that exceed
information--theoretic principles such as those suggested by the
CBH theorem, or by subjectivist Bayesian approaches to quantum
mechanics (see \eg Fuchs and Peres 2000, Fuchs 2002, Caves {\em et
al.} 2002, Spekkens 2004, and references therein). For
convenience, we shall frame our discussion in the context of the
Bayesian approach. After presenting our argument in the form of a
thought experiment, we shall make the link to Bub's principle
approach.

The Bayesian approach to quantum theory is based on an epistemic
attitude according to which the quantum state does not represent a
real physical state of a system, but instead supplies an observer
with statistical information concerning all possible distributions
of measurement results. The probabilities computed by the standard
Born rule are understood as probabilities of {\em finding} the
system on measurement in some specific state. Applying von
Neumann's projection postulate to the quantum state (or more
generally applying L\"uder's rule), under this account, is just an
adjustment of subjective probabilities, conditionalizing on newly
discovered results of measurement, \ie it is merely a change in
the observer's knowledge, or probability assignments. By contrast,
the unitary and linear quantum mechanical dynamics (\ie the
Schr\"odinger equation in the non-relativistic case) describes the
observer--independent and in this sense objective time evolution
of the quantum probabilities when no measurement takes place.
Hence, in this approach measurements can be treated operationally
as `black boxes' and require no further theoretical analysis.

\subsection{A thought experiment}\label{exp}
We now turn to our thought experiment which is a variant on Wigner's
friend. Consider the following set--up in which an observer $A$
measures the $z$--spin of a spin-half particle $P$ by means of a
Stern--Gerlach apparatus (which, to keep things simple, we omit
from our description below). The quantum state of $P+A$ initially
is
\begin{equation}
 \ket{\Psi_0}=(\alpha\ket{+_z}+\beta\ket{-_z})\ket{\psi_0}_A
\label{eq:initial}
\end{equation}
where $\abs{\alpha}^2+\abs{\beta}^2=1$ ($\alpha, \beta \not=0$),
the kets $\ket{\pm_z}$ are the $z-$spin eigenstates and $\ket{\psi_0}$ is the
initial {\em ready} state of $A$.
After the measurement, in a no collapse theory, the quantum mechanical state
of $P+A$ is the superposition:
\begin{equation}
\ket{\Psi_1}= \alpha\ket{+_z}\ket{{\rm see\; up}}_A+
\beta\ket{-_z}\ket{{\rm see\; down}}_A
\label{eq:final1}
\end{equation}
where $\ket{{\rm see\; up}}_A$ and $\ket{{\rm see\; down}}_A$ are,
say, the brain states of $A$ corresponding to her perceptions and
memories of the two possible outcomes of the measurement. By
contrast, in a collapse theory of the GRW kind, the state
(\ref{eq:final1}) is highly unstable (assuming that the chain of
interactions leading to $A$'s different memory states involves
macroscopically distinguishable position states), so that by the
time the measurement is complete this state collapses onto one of
its components.

Consider now an observable $\hat O$ of the composite system $P+A$
of which the state (\ref{eq:final1}) is an eigenstate with some
definite eigenvalue, say $+1$.\footnote{Observables like $\hat O$
are defined in the tensor product Hilbert space ${\H}_P\ten
{\H}_A$ unless superselection rules are introduced. For our
purposes think of $\hat O$ as an observable that pertains to $P$'s
spin degree of freedom and the relevant degrees of freedom of
$A$'s sense organs, perceptions, memory, etc.} Suppose that the
composite system $P+A$ is completely isolated from the
environment, and that a measurement of $\hat O$ is about to be
carried out on $P+A$ immediately after the state (\ref{eq:final1})
obtains. According to no collapse quantum mechanics the
measurement of $\hat O$, under these circumstances, is completely
non-disturbing in the sense that after the measurement the state
of $P+A$ remains precisely as in (\ref{eq:final1}). One may think
of $\hat O$ as an observable that is maximally sensitive to
whether or not the interference terms between the different
components of (\ref{eq:final1}) exist. In other words, the
measurement of $\hat O$ on $P+A$ if the state (\ref{eq:final1}) is
the true state of $P+A$ is a non-demolition measurement that, as
it were, passively {\em verifies} whether or not $P+A$ is in fact
in that state.

Note that $\hat O$ commutes neither with the $z-$spin nor with
$A$'s perceptions and memories of the outcomes of the $z-$spin
measurement. This surely raises interesting questions about the
status of the uncertainty relations in this set up and about the
reliability of $A$'s memories of the outcome of her spin
measurement in the event that $A$ measures $\hat O$ just after her
spin measurement. However, no matter what happens during the
measurement of $\hat O$ (to $A$'s memory of the outcome of her
spin measurement, or to the $z-$spin values themselves) quantum
mechanics implies that the {\em correlations} between the $z-$spin
of $P$ and $A$'s memories must remain exactly the same as they
were before the $\hat O$--measurement. Moreover, in a no collapse
theory, the state of $P+A$ immediately after the $\hat
O$--measurement will be, with complete certainty, just:
\begin{equation}
 \ket{\Psi_2}=\alpha\ket{+_z}\ket{{\rm see\; up}}\ket{{\rm see}\;
 \hat O=+1}+\beta\ket{-_z}\ket{{\rm see\; down}}\ket{{\rm
see}\;\hat O=+1} \label{eq:final2}
\end{equation}

where $\ket{{\rm see}\;\hat O=+1}$ is the state corresponding to
perceiving the result of the $\hat O$--measurement.

By contrast, in a collapse theory of the GRW kind, the state of
$P+A$ {\em immediately} after the $\hat O$--measurement will be
given by {\em one} of the eigenstates of $\hat O$, where the
probability that it will be state (\ref{eq:final1}), and therefore
that the outcome $\hat O=+1$ will be obtained, is only
$\abs{\alpha}^2$. Note that even if that outcome will obtain, the
state (\ref{eq:final1}) will extremely quickly collapse, again,
onto one of the components of state (\ref{eq:final1}) with
probabilities that are given by $\abs{\alpha}^2$ and
$\abs{\beta}^2$. So, in the GRW theory, the final value of the
$z-$spin and the spin memory of $A$ might be different before and
after the $\hat O$--measurement.

Note further that we deliberately do not specify here {\em who}
carries out the $\hat O$--measurement (\ie in which degree of
freedom the outcome $\hat O=+1$ is recorded). It may be carried
out by $A$ or by some other observer $B$ external to $A$'s
laboratory. As can be seen from our notation in (\ref{eq:final2}),
we have assumed that the outcomes of $A$'s spin measurement and of
the $\hat O$--measurement are recorded in {\em separate} degrees
of freedom. Quantum mechanics imposes no restrictions whatsoever
on the way in which the outcomes of these measurements are
recorded, except that they cannot be recorded {\em simultaneously}
in the same degree of freedom (since $\sigma_z\ten\id$ as well as
$A$'s memory observable are incompatible with $\hat O$). No
further restrictions are imposed by quantum mechanics (with or
without collapse) on the {\em identity} of the observers who may
carry out $\hat O$--type measurements (we return to this point
below).\footnote{See Albert (1983) for an extended discussion of
$\hat O$--type measurements and their implications. Aharonov and
Albert (1981) use $\hat O$--type measurements in their discussion
of the collapse of the quantum state in a relativistic setting.}

However, consider what would happen in our scenario if $A$ were to
communicate her outcome to $B$. As long as $B$ does not know the
outcome of $A$'s spin measurement, $B$'s predictions for the $\hat
O$ measurement will conform to option (b) (but see below). So if
$A$ and $B$ don't communicate, $B$'s predictions are {\em not}
ambiguous (unlike $A$'s predictions). What about $A$'s
predictions? Once $A$ communicates her spin outcome to $B$, $A$'s
predictions about the $\hat O$--measurement would no longer be
ambiguous, since the reduced state of $P+A$ after $A$'s
communication with $B$ would be given by the mixture corresponding
to the state (\ref{eq:final1}) (and the total state of $P+A+B$
would no longer be an eigenstate of $\hat O$). In fact, the
predictions of $A$ and $B$ would coincide in this case; that is,
both $A$ and $B$ would predict probability of $1/2$ to the outcome
$\hat O=+1$. However, quantum mechanics entails the existence of
{\em other} $\hat O$--type observables (\eg any observable $\hat
O_2$ of which the state of $P+A+B$ immediately after the
interaction of $A$ and $B$ is an eigenstate) with respect to which
we can run the {\em same} argument again, this time on the
predictions of both $A$ {\em and} $B$. And likewise {\em ad
infinitum}. But now the crucial point is that whether or not $A$
communicates her outcome to $B$ becomes irrelevant. Suppose that
$A$ doesn't.  Then, $A$ and $B$ differ in their predictions about
the outcome of the $\hat O$--measurement. And so one of them at
least must be wrong, or so it seems.

To make things simple, let us suppose that the $\hat
O$--measurement is to be carried out by the external observer $B$, but
let us consider $A$'s {\em predictions} of the
probabilities of the outcomes of the $\hat
O$--measurement.\footnote{Clearly, in quantum mechanics the
quantum state assigned to a system is supposed to give the
probabilities of the outcomes for {\em all} possible
measurements.} Here we encounter a problem in the
information--theoretic approach since it doesn't tell us on what
quantum state $A$ should base her predictions.
In order to calculate her expected probabilities $A$ may choose
one of the following two options.
\begin{enumerate}\item[(a)] Update her quantum state in
accordance with the outcome of the spin measurement she actually
observed, either $\ket{{\rm see\; up}}_A$ or $\ket{{\rm see\;
down}}_A$. In this case, she would collapse the state
(\ref{eq:final1}) onto one of its spin$+$memory components.
Applying the Born rule to this state, she will predict that the
result of the $\hat O$--measurement will be $+1$ with probability
$\abs{\alpha}^2$. \item[(b)] Ignore the outcome of the spin measurement
she actually observed, and conditionalize her probabilities on the
{\em uncollapsed} state as in (\ref{eq:final1}). In this case,
since the state in (\ref{eq:final1}) is an eigenstate of $\hat O$
with eigenvalue $+1$, she will predict that the result of the
$\hat O$--measurement will be $+1$ with {\em certainty} (\ie
probability $1$). \end{enumerate}
But by actually performimg a series of repeated $\hat
O$--measurements on identically prepared systems, all in state
(\ref{eq:final1}), we can in principle distinguish experimentally
between the two predictions. So, on pain of inconsisteny, no theory
can accept both predictions as true.

The point to be made here, however, is that the
information--theoretic approach gives no plausible account of {\em
which} option, (a) or (b), is the correct one. On the one hand, the full
information about the lab available to $A$ before the $\hat
O$--measurement is given by the {\em collapsed} state, and this implies
that option (a) is true. On the other hand if the dynamics of
quantum states is invariably given by the Schr\"odinger equation, then
(assuming that this information is admissible to $A$), her
predictions ought to be guided by the {\em uncollapsed} state
as in (\ref{eq:final1}). And so this means that option (b) is true.
And so we seem to have a straightforwrd inconsistency.

\begin{sloppypar}
We have formulated the above argument in terms of the Bayesian
information--theoretic approach. The link to Bub's principle
approach goes as follows. In both approaches quantum mechanical
measurements are construed {\em operationally} as `black boxes'
with no further analysis. This is precisely the sense intended by
Bub (2005, Sec. 4) of taking quantum information as a primitive
and irreducible physical concept. As argued by Bub, once we accept
the three constraints suggested by the CBH theorem, it follows
that measurements are to be treated operationally and measuring
apparatuses as black boxes. But, any `black box' theory of
measurement is bound to be incomplete in accounting for
Wigner's—type scenarios as spelled out above. Moreover, as long as
the mathematical formalism of quantum mechanics remains intact,
any operational approach that is not committed to either a
collapse--type theory or a no--collapse theory of a sort (on this
point, see also the next subsection) will necessarily run into
circumstances in which the assignment of quantum states will
become ambiguous, and as we just argued this ambiguity will lead
to inconsistent probability assignments to measurement outcomes!
If quantum theory is, indeed, on the right track (e.g., if there
no unknown superselection rules), then the decision between
collapse and no--collapse dynamics is forced upon us by {\em
empirical} constraints. In fact, we take this simply to mean that
the measurement problem in quantum mechanics by {\em constructive}
theories\footnote{This includes also the many worlds theory in all
its variants. Of course, one may interpret the NO BIT condition as
a dynamical constraint, which {\em a forteriori}, implies a
no--collapse theory. But then the issue of a principle approach
vs. a constructive one is off the table. See also footnote 18.}
\end{sloppypar}

\subsection{Some resolutions}\label{res}
In what follows we examine various possible objections to our
criticism (above).

{\bf (I)} {\bf $\hat O$--type measurements are not feasible.} As a
matter of fact, the $\hat O$ measurement is physically impossible
to carry out due to decoherence and the complexity of the set--up
under consideration. As far as our best neurophysiological
theories tell us, the sensory apparatus and brain processes of a
human observer involved in typical perception and memory processes
are {\em macroscopic} and subject to continuous decoherence
(induced by interactions in—and out—side of the brain) that cannot
be screened off. Consequently, $\hat O$ will become extremely fast
the wrong observable to measure in order to detect the
interference terms in states of the form of (\ref{eq:final1}).
Moreover, even if one could identify the right observable to
measure at a given time we cannot expect to have control over all
the relevant degrees of freedom in and outside the observer's
brain.

{\bf Reply (I)} Obviously, $\hat O$--type measurements are
extremely hard to carry out and need be continuously protected
against decoherence. This is much beyond our experimental reach.
In fact, even if we set decoherence aside, $\hat O$--type
measurements are not quite feasible in microscopic superpositions
over, say, only spin degrees of freedom (since we need to measure
total spin without measuring the spin components separately in
order for the measurement to be non-disturbing). But ways to
overcome such problems may be found. For example, in spin--echo
experiments we know today by means of macroscopic manipulations
only how to screen--off the effects of decoherence for appreciable
time intervals (see Hemmo and Shenker 2005). Moreover, feasibility
considerations are quite beside the point in the present context.
Quantum mechanics allows $\hat O$--type measurements, and the
above ambiguity must be resolved independently of whether we can
or cannot practically translate it into an experimental context.
Obviously, it may turn out that, on the basis of some new physics,
e.g., quantum gravity, the $\hat O$--measurement would be
impossible {\em in principle\/} on pain of violating certain new
physical laws, but as far as standard quantum theory is concerned,
this is not the case.

{\bf (II)} {\bf Option (b) is wrong.} $A$ should stick to option
(a) and use her spin--memory eigenstate in order to calculate her
probabilities (likewise for the external observer $B$). By
construction, the above set--up stipulates that $A$ carried out a
{\em measurement}, and therefore from $A$'s point of view the
state in (\ref{eq:final1}) has collapsed. As to $B$, although he
doesn't know which outcome $A$ has observed in the lab, he knows
that she has carried out a measurement. Therefore, he ought to
condition his predictions on the mixture of components in
(\ref{eq:final1}) rather than the full superposition. On this view
the predictions of $A$ and $B$ for the $\hat O$ measurement
coincide: they both give probability of one--half to $\hat O=+1$
despite the fact that their predictions are conditional on
different quantum states.

{\bf Reply (II)} The above argument is wrong--headed in the
context of the information--theoretic approach, since it seems to
imply that a measurement induces a stochastic transition from a
pure to a mixed state. This, however, is exactly what's argued by
collapse theories which aim at {\em constructing} a dynamical
theory in which such transitions can be accounted for. To put
matters differently, option (b) cannot in general be rejected
unless one is committed to a genuine collapse theory. If the
mathematical fomalism of quantum mechanics is not changed (\eg by
adding some suitable superselection rules), then given the
standard ways of thinking about it, it is a fact that no collapse
theories that employ dynamics which only induces effective
collapses (somehow within the global envelope of the unitary
dynamics of the quantum state) prescribe different predictions for
$\hat O$--type measurements than those of genuine collapse
theories, precisley because of the difference in their dynamical
laws for the quantum state.

{\bf (III)} {\bf Options (a) and (b) are compatible.} Which
option, (a) or (b), will turn out to give the right predictions
for the $\hat O$--measurement depends on {\em who} carries out the
measurement, $A$ or $B$. The outcome $+1$ of the $\hat
O$--measurement is assigned probability $1$ only if $A$ carries
out the measurement, but probability $1/2$ if $B$ carries out the
measurement. There is no inconsistency, because the different
probabilities are assigned on the basis of incompatible quantum
states (i.e. eigenstates of incompatible---maximally
non-commuting---observables) which cannot be identified with a
single system (or a single point of view). In order for the $\hat
O$--measurement to pick up the interference terms, and thus yield
the $+1$ outcome with probability $1$ in accordance with $B$'s
prediction, the apparatus must be kept completely isolated from
the content of $A$'s laboratory. If, before the $\hat
O$--measurement but just after $A$'s spin measurement, the
apparatus gets coupled to even a single air molecule or a photon
that interacted with the $z$--spin measuring device or with $A$'s
relevant degrees of freedom, its initial quantum state will
`split', and following the $\hat O$--measurement it will split
again relative to each component of the state (\ref{eq:final1})
exhibiting both $\pm 1$ outcomes (\ie the $\pm 1$ outcomes will be
probabilistic). It is highly plausible due to standard decoherence
effects (with a realistic Hamiltonian) that if $A$ carries out the
measurement herself she will have to interact with the $\hat
O$--apparatus, if only to switch it on, in just this way.
Therefore, there is a fundamental distinction between
$A$--systems, \ie systems that are coupled to the content of $A$'s
laboratory, and $B$--systems, \ie systems that are not so coupled,
and the probability distribution of the outcomes of the $\hat
O$--measurement depends on which kind of system actually carries
out the measurement.\footnote{Note that this dependence of the
actual probability distribution of the $\hat O$--measurement on
the $A$-- or $B$-- identity is absolutely crucial. Otherwise, the
probability assignments will be straightforwardly inconsistent, as
we argue above.}

{\bf Reply (III)} The above argument requires a constructive
notion of an observer or of measurement in order to make sense of
the fundamental distinction between the $A$-- and the $B$--
systems. In this sense it does not fit a ``black-box'' treatment
of measurement. Here are some problems that such a constructive
view will have to resolve. First, it is not at all clear what is
it that makes the $A-B$ distinction so fundamental? In other
words, How can a physically meaningful distinction between $A$--
and $B$-- systems be made on the basis of quantum mechanics as we
know it, or some other known physical principles? For example, $A$
might prepare {\em in advance} the $\hat O$--apparatus in its
ready state, and we can just assume that the $\hat O$--measurement
immediately and extremely quickly follows the spin measurement. In
that case, $A$ would be acting as a $B$--system, but she would
also {\em know} the outcome of the spin measurement. Second, why
can't our scenario be compatible with the case in which $A$ stores
her memory of the spin outcome in some perfectly {\em isolated}
degree of freedom in her brain (\ie her information is completely
private!), so that when performing the $\hat O$--measurement she
is actually acting as a $B$--system? It doesn't seem to us that
physical laws as we know them dictate that $A$'s information about
the $\hat O$--outcome need be public in the sense that leads to
decoherence. But if no restriction of this kind can be imposed
here, then the inconsistency isn't resolved: what would be in this
case $A$'s predictions about an upcoming $\hat O$--measurement?

{\bf (IV)} {\bf Option (a) is wrong.} Only the full quantum state
in (\ref{eq:final1}) is the right one for conditionalization (for
both $A$ and $B$), because the $\hat O$--measurement involves
interference between the two components of the state in
(\ref{eq:final1}). $A$'s information about the outcome of the spin
measurement becomes completely irrelevant due to the {\em
objective} (observer-independent) features of the dynamics of the
quantum state involved in the $\hat O$--measurement itself. One
might even argue that this is the `flipped-side' of the
uncertainty relations: since the value of $\hat O$ is known in
advance with certainty, $A$'s information about the spin outcome
is unreliable, and this is manifest by the fact that $A$'s
`memory' of the spin value (\ie her record observable) and $\hat
O$ are maximally non--commuting. Moreover, as suggested by Bub
(2005, Sec. 4), the emergence of classical information is
explained only by decoherence. But, by construction, the $\hat
O$--measurement requires {\em recoherence} of the components in
(\ref{eq:final1}). Therefore, whatever information these
components carry ought to be disregarded (by both $A$ and $B$) in
the face of the $\hat O$--measurement. Hence option (a) is no
option, and the above inconsistency is resolved.

{\bf Reply (IV)} This argument, again, implicitly relies on some
sort of a constructive theory of measurement. Consider the claim
that $A$'s memory of her spin outcome is {\em unreliable} in the
face of the $\hat O$--measurement. In so far as standard
(operational) quantum mechanics goes, we are faced with a
situation in which the measurement of spin on $P$ is followed by a
measurement of $\hat O$ on $P+M$. Since $\sigma_z\ten\id$ and
$\hat O$ are (maximally) non--commuting, by the uncertainty
relations, the outcomes of the spin measurement only determine the
probabilities of the outcomes of the $\hat O$--measurement (in
accordance with the Born rule). That's about all we can say in an
operational ``black box'' description of measurement. Whether or
not $A$'s memories are reliable {\em during} the measurement of
$\hat O$ is a question that cannot be settled by a quantum theory
that treats measurements as ``black boxes'' nor by the CBH
constrains on information transfer. It cannot even be settled by
appealing to the fact that perceptions and memories need be
modelled by decohering systems. The above argument requires
additional laws (over and above the laws of quantum mechanics)
about the dynamical behavior of $A$'s memory during the
interference involved in the $\hat O$--measurement.\footnote{In
Bohm's theory, for example, $A$'s memory of the spin is perfectly
reliable since the trajectories given by Bohm's deterministic
guidance equation cannot cross each other. By contrast, in modal
interpretations where the dynamics of the extra values is
stochastic, $A$'s memory of the spin might flip during the $\hat
O$--measurement. Anyway, on either theory the analysis of $\hat
O$--type measurements is a straightforward physical analysis, and
neither says that such measurements are {\em un}physical or cannot
be carried out, or what have you. There is a subtle issue
concerning $\hat O$--—type measurements in decoherence-based many
worlds theories, where the branching is defined by environmental
decoherence, since $\hat O$--type measurements involve recoherence
of the branches associated with A's memories of her spin
measurement. So it is not clear whether in our scenario the
branches associated with A's memories are well-defined, and
therefore whether or how these memories might play a role in A's
predictions (\eg of the outcome of the $\hat O$--measurement). But
we go no further on this issue here.} But, strictly speaking, such
laws are nothing but laws about the behavior of hidden (or extra)
variables of a sort. And so accepting (IV) above presupposes that
quantum mechanics is {\em incomplete}.

In the above discussion we've tried to counter various objections
to the $\hat O$--scenario. It is sometimes claimed that such a
scenario is, for some reason, excluded \textit{as a matter of
principle} by quantum mechanics. For example, it is claimed that a
single observer (say $A$) cannot carry out both the spin
measurement and the $\hat O$--measurement, since $\hat O$ is an
observable of $P+A$, and therefore measuring it by $A$ entails
some problematic form of {\em self-reference}. It is also claimed
that because $\hat O$ and $A$'s spin memories (as we said above)
don't commute, no single observer can ever be in a position to
know (with complete certainty) the values of both $\sigma_z$ and
$\hat O$ (as manifested by the uncertainty relations). So if $A$
knows the value of $\sigma_z$ (as we assume) she cannot possibly
know simultaneously also the value of $\hat O$. Both claims seems
to take it that quantum theory itself imposes some sort of
physical constraints as to the identity of the observers who may
carry out $\hat O$--type measurements, or that $\hat O$--type
measurements are for some reason meaningless. Therefore, the
inconsistency of $A$'s predictions is somehow unphysical or cannot
in principle be revealed by a single experiment.\footnote{That is,
it is claimed that our argument presupposes some sort of a
God's--eye view which, by quantum mechanics, is unavailable.}

However, these claims and various variants thereof are completely
off the mark in the context of our thought experiment. First, the
question of whether or not the $\hat O$-scenario involves
self--reference of any sort is {\em irrelevant\/}, since the issue
boils down in its entirety (as we presented it above) to a
straightforward question about the {\em predictions} of $A$ as to
the statistics of the outcomes of the $\hat O$--measurement---{\em
no matter who carries out the measurement\/}! Second, as a matter
of fact, our contraption above would involve no problematic form
of self-reference even if we were to assume that the $\hat
O$--measurement is to be carried out by $A$ herself.\footnote{This
is essentially because we could always assume that $A$'s memories
of the outcomes of the spin measurement and the $\hat
O$--measurement are to be associated with separate physical
features of, say $A$'s brain. So the scenario requires {\em no}
genuine self-reference.} Third, standard quantum mechanics imposes
no physical constraints whatsoever on the identity of the
observers who may carry out $\hat O$--type measurements. Fourth,
our argument above leads to no sort of infinite
regress.\footnote{For a more extensive discussion of some of these
issues, see Albert (1983, 1990).}

Taking stock, by this we have established our thesis III above,
namely, that the information-—theoretic approach is {\em
incomplete} and that this incompleteness leads to inconsistent
statistical predictions. Therefore, the approach must be supplemented by
further constructive laws that will remove the inconsistency. We see no
way of escaping the conclusion that $\hat O$--type
scenarios require some form of a constructive theory of
measurement. This is why we believe that neither
information--theoretic approaches nor ``black-boxes'' approaches
to quantum measurement circumvent Bell's (1990) objection that
decoherence is not enough in order to make sense of quantum
measurement---no matter whether or not `measurement' is
operationally construed.

\section{Conclusion: the Ancilla Argument}\label{anc}

In this paper we have criticized the information--—theoretic
approach to quantum mechanics and Bub's `principle' reading of the
CBH theorem. Agreed, the theorem remarkably demonstrates that
certain salient features of quantum mechanics as we know it (by
empirical observation) can be expressed very elegantly in
information--theoretic terms. But it does not support any
preference to the principle view of quantum theory over its
constructive counterparts. Moreover (and unlike other constructive
no collapse theories), our thought experiment clearly shows that
the information--theoretic approach does not address major
unresolved interpretational issues of quantum theory some of which
await {\em empirical} resolution.\footnote{See also Hagar 2003.}

More generally, the question raised by this paper boils down to
the following. Suppose that crucial experiments that are capable
of distinguishing between, say the GRW theory and environmentally
induced decoherence were to come out (no kidding!) in accordance
with the GRW predictions to a very good
approximation.\footnote{For recent progress see Adler 2005, Adler
{\em et al\/}. 2005, Bassi {\em et al\/}. 2005a. See also Hemmo
and Shenker 2005 for crucial experiments between collapse and no
collapse quantum theories testing thermodynamical effects.}
Nevertheless, it is often argued that the principle
(information--theoretic) approach to quantum mechanics would
remain {\em intact} for the following reason.

\begin{sloppypar}
Suppose that an {\em open} system $S$ is subjected to {\em
perfect} decoherence, namely to interactions with some degrees of
freedom in the environment $E$, such that the environment states
become strictly orthogonal. Suppose further that we have no access
whatsoever (as a matter of either physical fact or law) to these
degrees of freedom. In this case, the GRW dynamics for the density
operator of $S$ would be indistinguishable from the dynamics of
the {\em reduced} density operator of $S$ obtained by evolving the
composite quantum state of $S+E$ {\em unitarily} and tracing over
the inaccessible degrees of freedom of $E$. It turns out that this
feature is mathematically quite general, because the GRW dynamics
for the density operator is a completely positive linear map (see
Nielsen and Chuang 2000, pp. 353-373, and Simon, Buzek and Gisin
2001, especially fn. 14). From a physical point of view, this
means that the GRW theory is empirically equivalent to a quantum
mechanical theory with a unitary (and linear) dynamics of the
quantum state defined on a {\em larger} Hilbert space. In other
words, one could always introduce a new quantum mechanical {\em
ancilla} field whose degrees of freedom are inaccessible to us,
and cook up a unitary dynamics on the larger Hilbert space that
would simulate the GRW dynamics on the reduced density operator.
Therefore, experimental results that might seem to confirm
GRW-like dynamics could always be {\em re-interpreted} as
confirming quantum mechanical no collapse theories on larger
Hilbert spaces. In particular, such a theory could always be made
to satisfy the three CBH constraints, and thus save the principle
information--theoretic approach.
\end{sloppypar}

In so far as information--theoretic approaches are concerned it
seems to us that the above argument is quite premature. As the
above criticism of the ancilla argument is meant to show, it seems
to us that in the current state of quantum mechanics we are far
from being able to pinpoint a principle theory. In fact, we
believe that in general a principle approach is inferior relative
to a constructive one, whenever the latter is
possible.\footnote{This seems to be Einstein's later view despite
his introduction of special relativity as a principle theory; see
Brown and Timpson (2006) and references therein for an extended
account of this issue and of Einstein's views on this matter.} The
ancilla field in the above argument has, by construction, no
observable effects (see, e.g., Diosi 1989) and this amounts to
introducing extra variables (or more appropriately, a new `quantum
ether') into standard quantum mechanics, whose sole theoretical
role is to save some principles against (putative) empirical
refutation. If we are willing to accept {\em ad hoc} such an
argument in the context of non--relativistic quantum mechanics,
why should we reject similar `ether'--like approaches in the
context of relativity theory, or hidden variables theories in the
context of elementary quantum mechanics? One of the main reasons
that such latter approaches are sweepingly rejected (by
information theorists, among others) is that their complex
underlying structure doesn't translate into new empirical
predictions. But, by construction, the ancilla field in the above
argument doesn't translate into new empirical predictions either.
Moreover, although the ancilla theory could always be made to
satisfy the three CBH principles (in particular the NO BIT
condition) on the larger Hilbert space, unconditionally secure bit
commitment would in principle be possible (in the sense spelled
out in Section \ref{cp}, in so far as our experimental capacities
are concerned (as imagined in our scenario above), via protocols
that require Alice or Bob to access the ancilla field (which {\em
ex hypothesis} is inaccessible). So, the NO BIT principle as a
constraint on the feasible flow of information in the above story
becomes quite idle.

More generally, we accept that one {\em might} have good reasons
to protect unitary quantum mechanics against what would seem as
a straightforward empirical refutation. From a theoretical point of
view it might turn out that both collapse and hidden variables
theories could not be made compatible with some fundamental
physical principles which we cannot give up without giving up some
significant chunk of contemporary theoretical physics
(conservation of energy or Lorentz--covariance might be such
examples).\footnote{But compare Putnam (2005). We thank Itamar
Pitowsky for conversations about this issue.} In that case a
protective argument of the kind suggested above might be
understandable. But, in our view the current theoretical state of quantum
mechanics doesn't warrant such a stance. This is mainly because
although extremely successful, quantum mechanics itself has quite
deep foundational problems not only at its most basic level (\ie the measurement problem),
but also for example in its generalizations to {\em both} special and
general relativity. In such circumstances we believe that the
right epistemological stance is to suspend judgment and let
alternative theories, such as the GRW theory, flourish.\\ \\

\noindent \Large {\bf Acknowledgement} \\ \\
\normalsize We thank four anonymous referees, and Itamar Pitowsky
and Orly Shenker for very helpful comments. We especially thank
Guido Bacciagaluppi for reading an earlier draft of this paper and
for valuable and thoughtful suggestions and criticism. AH
acknowledges financial support from the Alexander von Humboldt
Foundation and the German Government through the Sofja Kovalskaja
Award through the PPM group in Konstanz.

\end{document}